\documentclass[12pt,a4paper]{article}
\newcommand\be[1]{\begin{equation} \label{#1}} 
\newcommand\ee{\end{equation}}
\newcommand\bea{\begin{eqnarray}}
\newcommand\eea{\end{eqnarray}}
\newcommand\ket[1]{|#1\rangle}
\newcommand\bra[1]{\langle #1|}

\newcommand\pty{{$\cal PT$}-symmetry}
\newcommand\ptc{{$\cal PT$}-symmetric}
\title{\pty\ and its spontaneous breakdown explained by
       anti-linearity}
\author{Stefan Weigert \\
HuMP (Hull Mathematical Physics)\\
              Department of Mathematics, University of Hull\\
              Cottingham Road, UK-Hull HU6 7RX,
              United Kingdom\\
\tt s.weigert@hull.ac.uk}
\date{September 2002}
\begin{document}
\maketitle
\begin{abstract}
The impact of an anti-unitary symmetry on the spectrum of
non-hermitean operators is studied. Wigner's normal form of an
anti-unitary operator is shown to account for the spectral
properties of non-hermitean, \ptc\ Hamiltonians. Both the
occurrence of single real or complex conjugate pairs of
eigenvalues follows from this theory. The corresponding energy
eigenstates span either one- or two-dimensional irreducible
representations of the symmetry ${\cal PT}$. In this framework,
the concept of a spontaneously broken ${\cal PT}$-symmetry is not
needed.
\end{abstract}
Deep in their hearts, many quantum physicists will renounce
hermiticity of operators only reluctantly. However, non-hermitean
Hamiltonians are applied successfully in nuclear physics, biology
and condensed matter, often modelling the interaction of a quantum
system with its environment in a phenomenological way. Since 1998,
non-hermitean Hamiltonians continue to attract interest from a
conceptual point of view \cite{bender+98}: surprisingly, the
eigenvalues of a one-dimensional harmonic oscillator Hamiltonian
remain {\em real} when the {\em complex} potential $\hat V =
i{\hat x}^3$ is added to it. Numerical, semiclassical, and
analytic evidence \cite{general} has been accumulated confirming
that bound states with {\em real} eigenvalues exist for the vast
class of {\em complex} potentials satisfying $V^\dagger(\hat x) =
V(- \hat x)$. In addition, pairs of complex conjugate eigenvalues
occur systematically.

\pty\ has been put forward to explain the observed energy spectra.
The Hamiltonian operators $\hat H$ under scrutiny are invariant
under the combined action of parity ${\cal P}$ and time reversal
${\cal T}$,
\be{ptinv}
[ \hat H , {\cal P T} ] = 0 \, .
 \ee
They act on the fundamental observables according to
\be{PTactions}
 {\cal P} : \left \{
 \begin{array}{l}
        \hat x \rightarrow - \hat x \, , \\
        \hat p \rightarrow - \hat p \, , \
 \end{array} \right. \qquad
 {\cal T} : \left \{
 \begin{array}{l}
       \hat x \rightarrow   \hat x \, , \\
       \hat p \rightarrow - \hat p \, , \
 \end{array} \right.
 \ee
and $\cal T$ {\em anti-commutes} with the imaginary unit,
\be{antilin}
{\cal T} i = i^* {\cal T} \equiv -i {\cal T} \, .
 \ee
Whenever a \ptc\ Hamiltonian has a {\em real} eigenvalue $E$, the
associated eigenstate $\ket{E}$ is found to be an eigenstate of
the symmetry ${\cal PT}$,
\be{realEptinv}
E=E^* : \quad \hat H \ket{E} = E \ket{E} \, , \quad
           {\cal PT} \ket{E} = + \ket{E}\, .
 \ee
Occasionally, ${\cal PT} \ket{E} = - \ket{E}$ occurs \cite{examples-1}
which is equivalent to (\ref{realEptinv}) upon redefining the
phase of the state: ${\cal PT} (i \ket{E}) = + (i \ket{E})$. There
is no difference between symmetry and anti-symmetry under ${\cal
PT}$.

However, if the eigenvalue $E$ is {\em complex}, the operator
${\cal PT}$ does {\em not} map the corresponding eigenstate of
$\hat H$ to itself,
\be{complEptnoninv}
E \neq E^* : \quad \hat H \ket{E} = E \ket{E} \, , \quad
                {\cal PT} \ket{E} \neq \lambda \ket{E}\; ,
                 \mbox{ any }\lambda \, .
 \ee
This situation is described as a `spontaneous breakdown' of \pty .
No mechanism has been identified which would explain this breaking
of the symmetry.

The \ptc\ square-well model provides a simple example for this
behavior \cite{znojil+01}. It describes a particle moving between
reflecting boundaries at $x=\pm 1$, in the presence of a piecewise
constant complex potential,
\be{squarepot}
V_Z(x) = \left \{
       \begin{array}{rl}
                     i Z \, ,& \quad x < 0 \, , \\
                    -i Z \, ,& \quad x > 0 \, ,
       \end{array} \right. \qquad Z \in I \! \! R \, .
\ee
Acceptable solutions of Schr\"odinger's equation must satisfy both
the boundary conditions, $\psi(\pm 1) =0$, and continuity
conditions at the origin. As long as the value of the parameter
$Z$ is below a critical value, $Z<Z_0^c$, the eigenvalues $E_n$ of
the non-hermitean Hamiltonian $\hat H = -\partial_{xx} + V_Z (x)$
are real, and each eigenstate $\ket{\psi_n}$ satisfies the
relations (\ref{realEptinv}), with eigenvalues $E_n$ and $+1$,
respectively. Above the threshold, $Z>Z_0^c$, at least one pair of
complex conjugate eigenvalues $E_0$ and $E_0^*$ develops. One of
the corresponding eigenstates has the form \cite{znojil+01}
\be{sqwfp1}
\psi_0 (x) =  \left \{
                    \begin{array}{rr}
                     K_p \sinh \kappa (1-x)\, , & \quad x > 0 \, ,\\
                     K_n \sinh \lambda^* (1+x)\, , & \quad  x < 0 \, ,
                     \end{array} \right.
\ee
the complex parameters $\kappa, \lambda, K_n,$ and $K_p$ being
determined by the boundary and continuity conditions. The state
$\psi_0 (x)$ is not invariant under ${\cal PT}$, i.e.
(\ref{complEptnoninv}) holds.

The purpose of the present contribution is a group-theoretical
analysis of \pty . The properties of \ptc\ systems are explained
in a natural way by taking into account that ${\cal PT}$ is not a
unitary but an {\em anti-unitary} symmetry of a {\em
non-hermitean} operator. The argument proceeds in three steps.
First, Wigner's normal form of anti-unitary operators is reviewed,
i.e. their (irreducible) representations are identified. Second,
the properties of non-hermitean operators with anti-unitary
symmetry are derived. These results are then shown to account for
the characteristic features of \ptc\ systems.

Wigner develops a normal form of anti-unitary operators $\hat A$
in \cite{wigner60}. Anti-unitarity of $\hat A$ is defined by the
relation
\be{anti-uni}
\bra{\hat A  \chi} \hat A \psi \rangle
   = \bra{\psi} \chi \rangle \,
 \ee
and it implies anti-linearity,
\be{antilin2}
\hat A (\alpha \ket{\psi} + \beta \ket{\chi})
     = \alpha^* \hat A\ket{\psi} + \beta^* \hat A \ket{\chi} \, .
\ee
which is equivalent to (\ref{antilin}). The representation theory
of $\hat A$ relies on the fact that the square of an anti-unitary
operator is {\em unitary}:
\be{square}
\bra{{\hat A}^2 \chi} {\hat A}^2 \psi \rangle
      = \bra{\hat A  \psi} \hat A \chi \rangle
      = \bra{\chi} \psi \rangle \, .
\ee
Therefore, the operator ${\hat A}^2$ has a complete, orthonormal
set of eigenvectors $\ket{\Omega}$ with eigenvalues $\Omega$ of
modulus one,
\be{A-squared}
{\hat A}^2 \ket{\Omega} = \Omega \ket{\Omega} \, , \quad
            | \Omega | = 1 \, .
\ee
It plays the role of a Casimir-type operator labelling different
representations of ${\hat A}$. Wigner distinguishes three
different types of representations corresponding to the
eigenvalues of ${\hat A}^2$: complex $\Omega \; (\neq \Omega)$,
$\Omega = +1 $, or $\Omega = -1$, summarized in Table
(\ref{irrepstable}).
\begin{enumerate}
\item An eigenstate $\ket{\Omega}$ of $\hat A^2$ with eigenvalue
$\Omega \; (\neq \Omega^*)$ is not invariant under $\hat A$.
Instead, the states $\ket{\Omega}$ and $\ket{\Omega^*} \equiv \hat
A \ket{\Omega}$ constitute a `flipping pair' with complex
`flipping value' $\omega$ (and $\omega^*$), where $\omega^2 =
\Omega$. They span a two-dimensional space which is closed under
the action of $\hat A$. Therefore, it carries a two-dimensional
representation of $\hat A$, denoted by $\Gamma_*$, which is {\em
irreducible}: due to the anti-linearity of $\hat A$, no (non-zero)
linear combination of the flipping states exist which is invariant
under $\hat A$.
\item Similarly, if ${\hat A}^2$ has an eigenvalue $\Omega =-1$,
then the operator $\hat A$ flips the states $\ket{-}$ and
$\ket{-^*} \equiv \hat A \ket{-}$. The flipping value is $i$, and
the associated two-dimensional representation $\Gamma_-$ is not
reducible.
\item Two different situations arise if there is an
eigenstate $\ket{1}$ of $\hat A^2$ with eigenvalue $+1$. The state
$\hat A \ket{1}$ is either a multiple of itself or not. In the
first case, the space spanned by $\ket{1}$ is invariant under
$\hat A$ and hence carries a one-dimensional representation
$\gamma_+$ of $\hat A$. When redefining the phase of the state
appropriately, one obtains an eigenstate $\ket{1}$ of $\hat A$
with eigenvalue $+1$. In the second case, the two states $\ket{+}
\equiv \ket{1}$ and $\ket{+^*} \equiv \hat A \ket{1}$ provide a
flipping pair with flipping value $\omega=+1$, and hence a
representation $\Gamma_+$. This representation, however, is {\em
reducible} due to the reality of the flipping value: the linear
combinations $\ket{1_r} = \ket{+} +\ket{+^*}$ and $\ket{1_i} = i(
\ket{+} -\ket{+^*})$ are both eigenstates of $\hat A$ with
eigenvalue $+1$.
\end{enumerate}
Consequently, a Hilbert space $\cal H$ naturally decomposes into a
direct product of invariant subspaces, each invariant under the
action of the anti-unitary operator $\hat A$,
\be{decompose}
{\cal H}= {\Gamma_*}^{\otimes N_*}
               \otimes   {\Gamma_-}^{\otimes N_-}
               \otimes {\Gamma_+}^{\otimes N_+}
               \otimes {\gamma_+}^{\otimes n_+} \, ;
\ee
the nonnegative integers $N_*$, $N_\pm$ and $n_+$ are related to
the degeneracies of the eigenvalues $\Omega \, (\neq \Omega^*)$
and $\Omega  = \pm 1$ of the operator ${\hat A}^2$. The
corresponding decomposition of a vector $\ket{\psi} \in {\cal H}$
is the closest analog of an expansion into the eigenstates of a
hermitean (or unitary) operator. Surprisingly, {\em
two}-dimensional irreducible representations of $\hat A$ exist
although there is only one generator, $\hat A$. No `good quantum
number' exists which would label the states spanning these
representations.

\begin{table}
  \centering
  \begin{tabular}{cclc}
        $\Omega \equiv \omega^2 $
           & $\Gamma$  & $\, \,$  action of ${\hat A}$ &  dim $\Gamma$ \\
               \hline \\
   $\Omega \neq \Omega^*$
          & $\Gamma_*$ &  $\begin{array}{l}
                  \hat A \ket{\Omega} = \omega^* \ket{\Omega^*} \\
                  \hat A \ket{\Omega^*} = \omega \ket{\Omega} \\
                \end{array}$   &  $2$   \\ \\
   $-1$  & $\Gamma_-$ &
   $\begin{array}{l}
     \hat A \ket{-}    = - i \ket{-^*} \\
     \hat A \ket{-^*} = +i \ket{-} \\
   \end{array}$
     &  $2$ \\ \\
 $+1$& $\Gamma_+$
 &  $\begin{array}{l}
           \hat A \ket{+} = + \ket{+^*}\\
           \hat A \ket{+^*} = + \ket{+}\\
       \end{array}$ &  $2$ \\ \\
$+1$ & $\gamma_+$
      &  $  \, \, \, \hat A \ket{1} = + \ket{1}$ & $1$ \\ \\
   \hline
  \end{tabular}
  \caption{Representations $\Gamma$ of the operator ${\hat A}$}
  \label{irrepstable}
\end{table}

A (diagonalizable) {\em non-hermitean} Hamiltonian $\hat H$ with a
discrete spectrum \cite{wong67} and its adjoint $\hat
H^\dagger$ each have a complete set of eigenstates:
\be{nonhermH}
\hat H \ket{\psi_n} = E_n \ket{\psi_n} \, , \quad \hat H^\dagger
\ket{\psi^n} = E^n \ket{\psi^n} \, ,
\ee
with complex conjugate eigenvalues related by $E^n = E_n^*$. They
form a {\em bi-orthonormal} basis in $\cal H$, as they provide two
resolutions of unity,
\be{unit}
\sum_n \ket{\psi^n}\bra{\psi_n}
 = \sum_n \ket{\psi_n}\bra{\psi^n}
 = \hat I \, ,
\ee
and satisfy orthogonality relations,
\be{dual}
\bra{\psi_m} \psi^n \rangle = \delta^n_m \, .
\ee

Let the non-hermitean operator $\hat H$ have an anti-unitary
symmetry $\hat A$,
\be{antiunitarysym}
 [ \hat H, \hat A] =0 \, .
\ee
Then the unitary operator $ \hat A^2$ commutes with $\hat H$, and
it has eigenvalues $\Omega$ of modulus one. Consequently, there
are simultaneous eigenstates $\ket{n,\Omega}$ of $\hat H$ and
$\hat A^2$:
\be{simultaneous*}
\hat H \ket{n,\Omega} = E_n \ket{n,\Omega} \, , \quad \hat A^2
\ket{n,\Omega} = \Omega \ket{n,\Omega} \, , \quad E_n \in {\sf C}
\, .
\ee
For simplicity, the eigenvalues $\Omega$ are assumed discrete and
not degenerate. Wigner's normal form of anti-unitary operators
suggests to consider three cases separately: complex $\Omega \,
(\neq \Omega^*)$ and $\Omega = \pm 1$.

$\underline{\Omega \neq \Omega^*}$ The state
\be{secondstate*}
\ket{n,\Omega^*} \equiv \omega \hat A \ket{n,\Omega} \, , \quad
             \omega^2=\Omega \, ,
\ee
is a second eigenstate of $\hat A^2$, with eigenvalue $\Omega^*$.
The states $\{ \ket{n,\Omega}, \ket{n,\Omega^*} \}$ provide a {\em
flipping pair} under the action of the operator $\hat A$,
\be{flip*}
 \hat A \ket{n, \Omega} = \omega^* \ket{n, \Omega^*} \, , \quad
 \hat A \ket{n, \Omega^*} = \omega \ket{n, \Omega} \, ,
\ee
carrying the representation $\Gamma_*$. No degeneracy of the
eigenvalue $E_n$ is implied by the anti-unitary $\hat A$-symmetry
of $\hat H$. However, the non-hermitean Hamiltonian has a second
eigenstate $\ket{n, \Omega^*}$ with eigenvalue $E_n^*$,
\be{implication*}
 \hat H \ket{n, \Omega^*} = E_n^* \ket{n, \Omega^*} \, ,
\ee
as follows from multiplying the first equation of
(\ref{simultaneous*}) with $\hat A$ and $\omega$.

$\underline{\Omega =-1}$ Formally, the results for the
representation $\Gamma_-$ are obtained from the previous case by
setting $\omega = i$. Again, a pair of complex conjugate
eigenvalues is found, and the associated flipping pair spans a
two-dimensional representation space.

$\underline{\Omega=+1}$ This case is conceptually different from
the previous ones as two possibilities arise. Consider the state
$\ket{n,+}$, an eigenvector of both $\hat H$ and $\hat A^2$ with
eigenvalues $E_n$ and $+1$, respectively. It satisfies Eqs.
(\ref{simultaneous*}) with $\Omega \to +$. If, on the one hand,
applying $\hat A$ to $\ket{n,+}$ results in $e^{i\phi} \ket{n,+}$,
then the state $\ket{n,1} \equiv e^{-i\phi/2}\ket{n,+}$ is an
eigenstate of $\hat A$ with eigenvalue $+1$,
\be{implic+1}
\hat A \ket{n,1} = \ket{n,1} \, .
\ee
This occurrence of the one-dimensional representation $\gamma_+$
forces the associated eigenvalue $E_n$ of $\hat H$ to be real
since
\be{implic+1b}
E_n \ket{n,1} = \hat H \hat A \ket{n,1} = \hat A \hat H \ket{n,1}
              = E_n^* \ket{n,1} \, .
\ee
If, on the other hand, $\ket{n,+^*} \equiv \hat A \ket{n,+}$ is
{\em not} a multiple of $\ket{n,+}$, then these states combine to
form the representation $\Gamma_+$, the flipping value being $+1$.
Further, the state $\ket{n,+^*}$ is an eigenstate of the
Hamiltonian with eigenvalue $E_n^*$. As the flipping number is
real, linear combinations of $\ket{n,+}$ and $\ket{n,+^*}$ do
exist which are eigenstates of $\hat A$---however, they are not
eigenstates of $\hat H$. Consequently, the anti-unitary symmetry
of the Hamiltonian makes itself felt (on a subspace with $({\cal
PT})^2 = +\hat I$) by either a single real eigenvalue or a pair of
two complex conjugate eigenvalues.

If any of the two-dimensional representations $\Gamma_*$ or
$\Gamma_\pm$ occurs and the associated eigenvalue happens to be
real, the anti-unitary symmetry implies a twofold degeneracy of
the energy eigenvalue. Again, the symmetry provides no additional
label, and simultaneous eigenstates of $\hat H$ and $\hat A$ can
be constructed for $\Gamma_+$ only. These cases will be denoted by
$\Gamma_*^d$ or $\Gamma_\pm^d$.

It will be shown now that the properties of ${\cal PT}$-symmetric
quantum systems are consistent with the representation theory of
non-hermitean Hamiltonians possessing an anti-unitary symmetry.
Upon identifying
\be{ptA}
\hat A = {\cal PT} \, ,
\ee
one needs to check the value of $({\cal PT})^2$ when applied to
eigenstates of the Hamiltonian in order to decide which of the
representations $\Gamma_*$, $\Gamma_\pm$, or $\gamma_+$, is
realized. Various explicit examples will be given now.

For parameters $Z<Z_0^c$, the eigenvalues of the \ptc\ square-well
are real throughout, and the operators $\hat H$ and ${\cal PT}$
have common eigenstates. Thus, the relations (\ref{realEptinv})
correspond to a multiple occurrence of the representation
$\gamma_+$, compatible with $({\cal PT})^2 = + \hat I$.

For $Z>Z_0^C$, the energy eigenstate $\psi_0(x) \equiv \bra{x} E_0
, + \rangle$ in (\ref{sqwfp1}) satisfies $({\cal PT})^2
\ket{E_0,+}$ $=$ $ + \ket{E_0,+}$. Therefore, the states
$\ket{E_0,+}$ and $\ket{E_0,+^*} \equiv {\cal PT}\ket{E_0,+}$
carry a representation $\Gamma_+$, and the presence of two complex
energy eigenvalues, $E_0$ and $E_0^*$ is justified. Eqs.
(\ref{complEptnoninv}) can be completed to read:
\be{complEptnoninv2}
E \neq E^* : \quad
 \begin{array}{l}
 \hat H \ket{E_0,+}   = E_0 \ket{E_0,+} \, ,\\
 \hat H \ket{E_0,+^*} = E_0^* \ket{E_0, +^*} \, ,
\end{array}
 \begin{array}{l}
 {\cal PT} \ket{E_0,+} = + \ket{E_0,+^*}\, , \\
 {\cal PT} \ket{E_0,^*} = + \ket{E_0,+} \, .
\end{array}
 \ee
Consequently, \pty\ is not broken but at $Z=Z_0^c$ the system
switches between the representations $\Gamma_+$ and $\gamma_+$,
with a corresponding change of the energy spectrum.

The following examples are taken from a discrete family of
non-hermitean operators \cite{khare+00},
\be{kharemandal}
{\hat H}_M = \hat p^2 - (\zeta \cosh 2x -i M)^2 \, , \quad \zeta
\in I \!\! R \, ,
\ee
$M$ taking positive integer values. Each operator ${\hat H}_M$ is
invariant under the combined action of ${\cal PT}$ where ${\cal
P}$ is parity about the point $a=i\pi/2$: $x \to i\pi/2 - x$. Due
to the reflection about a point off the real axis, the operators
${\cal P}$ and ${\cal T}$ do not commute as has been pointed out
in \cite{bagchi+01}. However, this fact is not essential here
since only the anti-unitary character of the symmetry ${\cal PT}$
is essential.

For $M=2$, two complex conjugate eigenvalues $E_+$ and $E_- =
E_+^*$ of ${\hat H}_2$ exist, with associated eigenstates
\be{khareM2}
\psi_+ (x) = \Psi(x) \cosh x  \equiv \bra{x} E_+,- \rangle \, ,
\quad \psi_- (x) = \Psi(x) \sinh x \equiv \bra{x} E_+,-^* \rangle
\, ,
\ee
and a ${\cal PT}$-invariant function $\Psi (x) = \exp [(i/2) \zeta
\cosh 2x ]$. These states are a flipping pair with flipping value
$i$,
\be{kharepair}
{\cal PT} \psi_+ (x) = -i \psi_- (x) \, , \quad {\cal PT} \psi_-
(x)  =   i \psi_+ (x) \, ,
\ee
and the twofold application of ${\cal PT}$ gives $(-1)$. Hence,
the representation $\Gamma_-$ is realized. Similarly, for $M=4$,
four eigenstates form two flipping pairs, i.e. two representations
$\Gamma_-$, each being associated with a pair of complex conjugate
eigenvalues.

For $M=3$, three different real eigenvalues of the Hamiltonian
${\hat H}_3$ have been obtained analytically if $\zeta^2 < 1/4$.
The corresponding eigenfunctions are given by
\be{khareM=3}
\psi(x) = \Psi(x) \sinh 2x \, , \quad \psi_\pm (x) = \Psi(x)
(A\cosh 2x \pm i B) \, ,
\ee
with real coefficients $A$ and $B$. Under the action of ${\cal
PT}$, the state $\psi (x)$ is mapped to itself, while $\psi_\pm
(x) $ each acquire an additional minus sign. Therefore, the states
$\psi (x)\equiv \bra{x} E,+ \rangle$ and $i\psi_\pm(x) \equiv
\bra{x} E_\pm \rangle $ are simultaneous eigenstates of $\hat H$
and ${\cal PT}$ with eigenvalues $+1$. The part of Hilbert space
spanned by these three states transforms according to three copies
of the representation $\gamma_+$. If $\zeta = 1/2$, the
eigenvalues $E_\pm$ turn degenerate, and the eigenstates given in
(\ref{khareM=3}) merge, $i\psi_+ (x) = i\psi_- (x) \equiv
\varphi(x)$. However, a second, independent ${\cal PT}$-invariant
solution of Schr\"odinger's equation can be found,
\be{secondM=3}
\phi (x) = \Psi (x) \int_{x_0}^x dy \frac{e^{-i\varphi
(y)/2}}{\varphi^2 (y)}  \, .
\ee
The solutions $\{ \varphi, \phi\}$ transform according to
$\gamma_+ \otimes \gamma_+ \equiv \Gamma_+^d $. So far, the
representation $\Gamma_*$ has apparently not been realized in
\ptc\ quantum systems---a possible explanation is the constraint
${\cal T}^2 = \pm 1$ for time reversal \cite{boehm79}.

In summary, the representation theory of anti-unitary symmetries
of non-hermi\-te\-an `Hamiltonians' has been developed on the
basis of Wigner's normal form of anti-unitary operators.
Typically, energy eigenvalues come in complex conjugate pairs, and
the associated eigenstates of the Hamiltonian span a
two-dimensional space carrying one of the two-dimensional
representations $\Gamma_*$, or $\Gamma_\pm$. Furthermore, a single
real eigenvalue may occur, related to a one-dimensional
representation $\gamma_+$. In this case a single $\hat
A$-invariant energy eigenstate state exists while there are no
simultaneous eigenstates of the Hamiltonian and the symmetry
operator in the two-dimensional $\hat A$-invariant subspaces.
Instead, flipping pairs of states can be identified. Generally,
the symmetry does not imply the existence of degenerate
eigenvalues---only if the Hamiltonian happens to have a real
eigenvalue, a two-dimensional degenerate subspace may exist
occasionally. These results naturally explain the properties of
eigenstates and eigenvalues of ${\cal PT}$-symmetric quantum
systems. In particular, it is not necessary to invoke the concept
of a {\em spontaneously broken} ${\cal PT}$-symmetry. Contrary to
a unitary or hermitean symmetry, the presence of an anti-unitary
symmetry does not imply the existence of a set of simultaneous
eigenstates of $\hat H$ and ${\cal PT}$---simply because an
anti-linear operator is not guaranteed to have a complete set of
eigenstates. Finally, the present approach provides a new
perspective on the suggested modification of the scalar product in
Hilbert space \cite{znojil01} which will be presented elsewhere
\cite{weigert02} in detail.

\end{document}